\documentclass[showpacs,amsmath,amsart,twocolumn,showkeys,pra,superscriptaddress]{revtex4-1}
\usepackage{dcolumn}    % Align table columns on decimal point
\usepackage{ifpdf}
\usepackage{amssymb,lineno,amsfonts}
\usepackage{graphicx}   % Include figure files
\usepackage{dcolumn}    % Align table columns on decimal point
\usepackage{bm}         % bold math
\usepackage{bbm}
\usepackage{mathrsfs}
\usepackage{upgreek}
\usepackage{mathtools}
\usepackage{epstopdf}
\usepackage{setspace}
\usepackage{hyperref}
\usepackage{float}
\usepackage{natbib}
\usepackage[usenames,dvipsnames]{xcolor}
\usepackage[matrix,frame,arrow]{xypic}
\usepackage{multirow}

\newcommand{\ket}[1]{\vert{#1}\rangle}

\newcommand{\outpr}[2]{\vert{#1}\rangle\langle{#2}\vert}

\newcommand{\proj}[1]{\vert{#1}\rangle\langle{#1}\vert}
\newcommand\at[2]{\left.#1\right|_{#2}}

\definecolor{med-blue}{RGB}{25,25,112}
\hypersetup{colorlinks, linkcolor={red},citecolor={blue}, urlcolor={MidnightBlue}}

\begin{document}
\title{Ancilla Induced Amplification of Quantum Fisher Information}
\author{C. S. Sudheer Kumar}
\email{sudheer.kumar@students.iiserpune.ac.in}
\affiliation{Department of Physics and NMR Research Center}
\author{T. S. Mahesh}
\email{mahesh.ts@iiserpune.ac.in}
\affiliation{Department of Physics and NMR Research Center}
\affiliation{Center for Energy Sciences, \\
	Indian Institute of Science Education and Research, Pune 411008, India}

\begin{abstract}
Given a quantum state with an unknown parameter, the Quantum Fisher Information (QFI) is a measure of the amount of information that an observable can extract about the parameter. QFI also quantifies the maximum achievable precision in estimating the parameter with a given amount of resource via an inequality known as quantum Cramer-Rao bound. In this work, we describe a protocol to amplify QFI of a single target qubit precorrelated with a set of ancillary qubits.
A single quadrature measurement of only ancillary qubits suffices to perform the complete quantum state tomography (QST) of the target qubit.
We experimentally demonstrate this protocol using an NMR system consisting of a $^{13}$C nuclear spin as the target qubit and three $^1$H nuclear spins as ancillary qubits. We prepare the target qubit in various initial states, perform QST, and estimate the amplification of QFI in each case. We also show that the QFI-amplification scales linearly with the number of ancillary qubits and quadratically with their purity.
\end{abstract}

\maketitle

\section{Introduction}
Quantum devices are expected to bring out a revolution in the way information is stored, manipulated, and communicated \cite{quant_info_neilson_chuang}.  An important criterion to achieve this goal is the capability to efficiently measure two-level quantum systems, or qubits \cite{Devincenzo}. Spin-based systems are  among various architectures which are being pursued for the physical realization of a quantum processor \cite{Laflam_quantcomp_review}.  Nuclear spins in favorable atomic or molecular systems have the capability to store quantum information for sufficiently long durations and to allow precise implementation of desired quantum dynamics.  Accordingly, Nuclear Magnetic Resonance (NMR) is often considered as a convenient testbed for  quantum emulations \cite{Corynmr_1st, cory, NMR_review_Cory}.  
In a conventional NMR scheme, tiny nuclear polarizations demand a collective ensemble measurement of about $10^{15}$ identical spin-systems.
There have been several proposals to increase the sensitivity of nuclear spin detection.
For example, dynamic nuclear polarization (DNP) transfers polarization from electrons to nuclei, thereby enhancing the nuclear polarization by 2 to 3 orders of magnitude \cite{doi:10.1063/1.2833582}.  
Optical polarization and detection often enables single-spin measurements, such as in the case of nitrogen vacancy centers in diamond \cite{wrachtrup1997magnetic}. 
Further improvements in sensitivity are possible by using quantum metrology  which has recently attracted significant research interests \cite{Metro_QFI_review}.  
 Cappellaro et. al. have proposed a metrology scheme by measuring a set of ancillary qubits after correlating them with the target qubit \cite{metro_Cory_entassisted}.
 $N$-spin quantum metrology in the presence of decoherence has been discussed by Knysh et. al. \cite{Metro_Nspin_decoherence}. Quantum metrology in a solid state NMR system exploiting spin-diffusion has been proposed by Negoro et. al. \cite{Metro_diffusn_Nspin}.

\begin{figure}
	\includegraphics[width=8cm,clip = true,trim = 0cm 0cm 0cm 0cm]{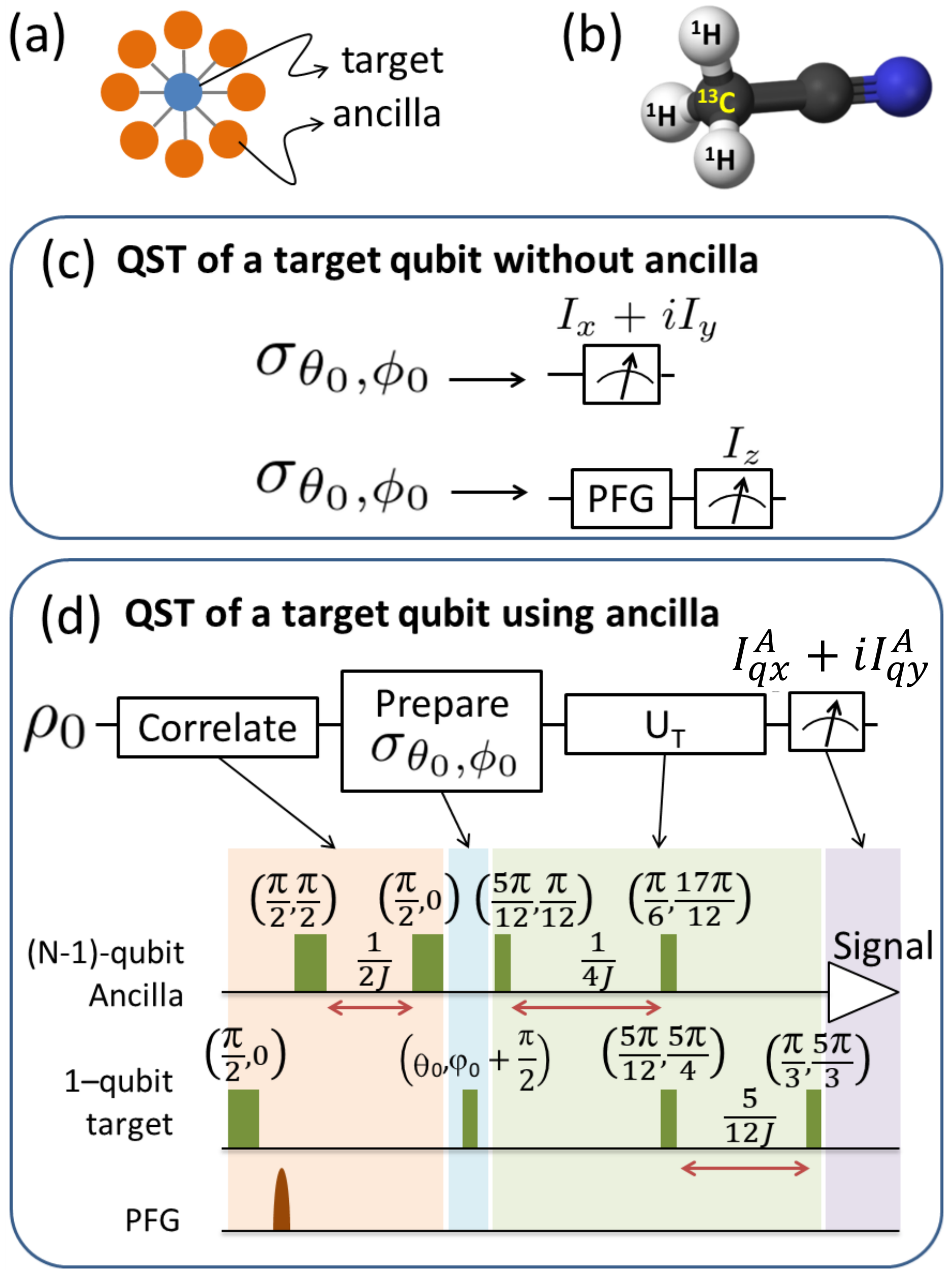}
	\caption{(Color online) (a) Schematic representation of an STR, (b) molecular structure of acetonitrile corresponding to a 4-qubit STR, (c) QST of a target qubit without ancilla, requiring two independent NMR experiments, (d) QST of a target qubit using ancilla, requiring a single quadrature detection of ancillary qubits without decoupling the target during acquisition. In (d), each RF pulse shown by a rectangle is labeled with two parameters - nutation angle and  phase respectively.  The tomography parameters are optimized using a genetic algorithm subject to certain constraints such as rank, condition number, and overall signal enhancement \cite{Abhishk_Ancla_asst_tomo}.}
	\label{Pulse_seq_fig}
\end{figure}

The present work involves a single target qubit and a set of ancillary qubits. While the methods described in the following are general, and can be adopted for a quantum register with a general topology, we particularly focus on star-topology registers (STRs).
An STR consists of a central target qubit uniformly interacting with a set of identical ancillary qubits which do not interact among themselves (see Fig. \ref{Pulse_seq_fig}(a)).  
Recently STRs have been utilized for several interesting applications.  
The main advantage of an STR is that it allows simultaneous implementation of C-NOT operations on the ancillary qubits controlled by the target qubit without requiring individual control of ancillary qubits.   Simmons et. al. exploited this property to prepare large NOON states and used them to sense ultra-low magnetic fields \cite{Metro_Jones_magntc_sensor}.  Abhishek et. al. proposed efficient measurement of translational diffusion in liquid ensembles of STR molecules \cite{Abhishek_NOON_diffusn}.  Using a 37-qubit STR, Varad et. al. demonstrated a strong algorithmic cooling of the target qubit by repeatedly releasing its entropy to the ancillary qubits \cite{Varad_Algocool}.  Deepak et. al. transferred the large polarization of the ancillary qubits directly to the long-lived singlet-state of a central pair of qubits in an STR-like register \cite{Deepak_star_singlet}.  More recently, Soham et. al. have utilized STRs to investigate the rigidity of temporal order in periodically driven systems \cite{Soham_timecrystal}. 

In this work, we propose and experimentally demonstrate a protocol to perform the full quantum state tomography (QST) of a target qubit in an STR. We find that a single-scan quadrature measurement of ancillary qubits of an STR precorrelated with the central target qubit is sufficient to tomograph the target qubit.  Moreover, this procedure leads to a strong amplification of Quantum Fisher Information (QFI) i.e., QFI scales linearly with the number of ancillary qubits and quadratically with their purity (for small purity, $\varepsilon_{a,1}\ll 1$). QFI quantifies the amount of information that a given observable can extract about a parameter of a quantum system in an unknown state  \cite{Metro_QFI_review}. Moreover, QFI allows one to estimate the quantum Cramer-Rao bound, which sets an upper-bound for the  achievable precision in estimating an unknown parameter with a given amount of resource \cite{Metro_kolodyski_QFI}.

In the following we first discuss the theoretical aspects of QST and QFI. In Sec. \ref{qst1q} we describe QST of the target qubit without using ancillary qubits.
In Sec. \ref{Tomo_with_precorltn} we describe QST of the target qubit after precorrelating it with ancillary qubits. In Sec. \ref{QFI_subsec} we discuss QFI corresponding to polar, azimuthal, and dual parameters of single (uncorrelated) as well as STR (correlated) systems.  In Sec. \ref{Expt_sectn} we describe experimental aspects of QST and estimation of QFIs.  Finally we summarize and conclude in Sec. \ref{Sumry_conclusn}.

\section{Theory}
\subsection{QST of a target qubit without ancilla}\label{qst1q}
Consider a single target qubit in a mixed state with a purity factor $\varepsilon_{t,1} \in[0,1]$.  In the Bloch sphere, we may represent it as a convex sum of the maximally mixed state $\mathbbm{1}_2/2$ and a surface point 
\begin{eqnarray}
\ket{\psi_{\theta_0,\phi_0}}=\cos(\theta_0/2)\ket{0}+ e^{i\phi_0} \sin(\theta_0/2)\ket{1}
\end{eqnarray}
so that the density matrix
\begin{eqnarray}
\varrho_{\theta_0,\phi_0} &= &(1-\varepsilon_{t,1})\mathbbm{1}_2/2 + \varepsilon_{t,1} \proj{\psi_{\theta_0,\phi_0}}\nonumber\\
&=&\mathbbm{1}_2/2+\varepsilon_{t,1} \sigma_{\theta_0,\phi_0}/2, \nonumber \\
&=&
\frac{1}{2}
\left[
\begin{array}{cc}
1+\varepsilon_{t,1}\cos\theta_0 & \varepsilon_{t,1} e^{-i\phi_0}\sin\theta_0  \\
\varepsilon_{t,1} e^{i\phi_0}\sin\theta_0  & 1-\varepsilon_{t,1}\cos\theta_0
\end{array}
\right],
\label{varrhoth0ph0}
\end{eqnarray}
where
\begin{eqnarray}
\sigma_{\theta_0,\phi_0} &=&\sin\theta_0\cos\phi_0~ \sigma_{x}+\sin\theta_0\sin\phi_0~ \sigma_{y}+\cos\theta_0 ~\sigma_{z}\nonumber\\
&=&\hat{n}_0.\vec{\sigma}.
\label{sigthph}
\end{eqnarray}
QST to determine the deviation part $\tilde{\sigma}_{\theta_0,\phi_0}$ of the experimental density matrix  can now be achieved using two independent experiments \cite{Abhishk_Ancla_asst_tomo} (see Fig. \ref{Pulse_seq_fig}(c)): (i) estimating $\phi_0$ via a quadrature measurement of $I_{x}+iI_{y}$, where $I_\alpha$ are the components of spin-angular momentum operators;
(ii) estimating $\theta_0$ via $I_z$ measurement after dephasing the off-diagonal terms using pulsed field gradient (PFG). 
The correlation  \cite{Correltn_fidlty_cory} between the expected ($\sigma_{\theta_0,\phi_0}$), and
the experimental ($\tilde{\sigma}_{\theta_0,\phi_0}$) deviation density matrices are calculated using
\begin{eqnarray}
C = \frac{\mathrm{Tr}[\tilde{\sigma}_{\theta_0,\phi_0}\sigma_{\theta_0,\phi_0}]}{\sqrt{\mathrm{Tr}[\tilde{\sigma}_{\theta_0,\phi_0}^2]\mathrm{Tr}[\sigma_{\theta_0,\phi_0}^2]}}.
\label{C_corelatn}
\end{eqnarray}

\subsection{QST of a target qubit in an STR}\label{Tomo_with_precorltn}
Here we consider an $N$-qubit STR consisting of a single target qubit surrounded by a set of $N-1$ indistinguishable ancillary qubits.  Under the weak-coupling approximation, Hamiltonian for the STR is of the form
\begin{eqnarray}
H = -\hbar\omega_t I_{1z} - \hbar\omega_a \sum_{j=2}^{N} I_{jz}+ \pi \hbar J \sum_{j=2}^{N} 2I_{1z}I_{jz},
\end{eqnarray}
where $\omega_t$ and $\omega_a$ are the resonance offsets of the target and ancilla respectively, $I_{jz}$ is the $z$-component of the spin-angular momentum operator corresponding to $j^{\mathrm{th}}$ qubit.  Because of the magnetic-equivalence symmetry,  the scalar couplings between the target and ancilla are all same, say of magnitude $J$, while those among the ancillary qubits become unobservable (Fig. \ref{Pulse_seq_fig} (a)).  In the following we describe (i) precorrelating a target qubit with ancillary qubits, (ii) implementing an arbitrary transformation on the target qubit, and (iii) QST of the target qubit's state by a single-shot quadrature measurement of the ancillary qubits (Fig. \ref{Pulse_seq_fig} (d)).

(i) In a strong Zeeman field $B_0$, 
the thermal equilibrium state of STR
under high-temperature approximation is of the form
\begin{eqnarray}
\rho_0 = \mathbbm{1}_{2^N}/2^N + \varepsilon_{t,N} I_{1z}+\varepsilon_{a,N}\sum\limits_{i=2}^{N}I_{iz}
\label{rho0}
\end{eqnarray}
where 
\begin{eqnarray}
\varepsilon_{t,N} = \frac{\hbar\gamma_t B_0}{2^Nk_BT} ~~\mbox{and}~~ \varepsilon_{a,N} = \frac{\hbar\gamma_a B_0}{2^Nk_BT}
\end{eqnarray}
are the purity factors (each $\lesssim 10^{-5}$) \cite{cavanagh}.  Here $\gamma_t$ and $\gamma_a$ are gyromagnetic ratios of the target and ancilla qubits respectively, and $k_B$ is the Boltzmann constant, $T$ is the absolute temperature. Thus the thermal state is practically an uncorrelated state.  We utilize the standard NMR technique, namely INEPT \cite{cavanagh} to prepare a correlated state of the form
\begin{eqnarray}
\rho_1 =\mathbbm{1}_{2^N}/2^N+\varepsilon_{a,N} 2 I_{1z}  \sum\limits_{i=2}^{N} I_{iz}.
\label{rho1}
\end{eqnarray} 
The corresponding pulse sequence is shown in Fig. \ref{Pulse_seq_fig}(d). For a large STR with  $\varepsilon_{a,N} > \varepsilon_{t,N}$, the above state corresponds to a large anti-phase spin-order  and accordingly leads to a strong NMR signal.  

(ii) The target is now ready for an arbitrary local transformation (unitary or nonunitary) so that the combined state takes the form,
\begin{eqnarray}
\rho_{\theta_0,\phi_0} = \mathbbm{1}_{2^N}/2^N+\varepsilon_{a,N} (\sigma_{\theta_0,\phi_0} \otimes \mathbbm{1}_{2^{N-1}}) \sum\limits_{i=2}^{N} I_{iz}.
\label{rho2}
\end{eqnarray} 
Here the deviation part characterizing the target qubit $\sigma_{\theta_0,\phi_0}$ (Eq. (\ref{sigthph})), with $\theta_0$ and $\phi_0$ being the unknown parameters, is to be determined via QST.

(iii) The task of QST can be transformed into solving a set of linear constraint equations \cite{Abhishk_Ancla_asst_tomo}.  While there is no unique solution to this task, one looks for a solution that optimizes certain parameters.  A numerical solution that maximizes the norm of the constraint matrix while simultaneously minimizing its condition number is described in $U_T$ part of Fig. \ref{Pulse_seq_fig} (d).  The state of STR after applying the tomography unitary $U_T$ is
$U_T \rho_{\theta_0,\phi_0} U_T^\dagger$.

Corresponding to the two Zeeman eigenstates $q \in \{0,1\}$ of the target qubit, and $\alpha \in \{x,y\}$, we define four transverse observables for the ancilla:
\begin{eqnarray}
M_{q\alpha} = U_T^\dagger I^A_{q\alpha} U_T, 
\label{Mqalpha}
\end{eqnarray}
where
\begin{eqnarray}
I^A_{q\alpha} =\left[(\proj{q} \otimes \mathbbm{1}_{2^{N-1}}) \sum_{j=2}^N I_{j\alpha} \right].
\label{Iaqalpha}
\end{eqnarray}
Thus 
\begin{eqnarray}
M_Q = \sum_q M_{qx}+iM_{qy}
\label{mqstr}
\end{eqnarray}
is the effective quadrature observable.
The expectation values of these observables are measured via the intensities
$ \mathrm{Tr}\left[\rho_{\theta_0,\phi_0} 
M_{q\alpha} \right]$
of corresponding NMR spectral lines separated by the coupling constant $J$.
Thus, a single-shot quadrature read-out of ancillary qubits provides four real constraints sufficient to determine the two unknowns $\theta_0$ and $\phi_0$, and hence achieve QST of the target qubit \cite{Abhishk_Ancla_asst_tomo}.

\subsection{Quantum Fisher Information}\label{QFI_subsec}
Consider a quantum system prepared in a state in the neighborhood of $\rho_{\theta_0,\phi_0}$ and
$M$ be a given observable with spectral decomposition $M=\sum_im_i\outpr{m_i}{m_i}$.    Let us first assume that the polar angle $\theta$ has a distribution around $\theta_0$ while $\phi_0$ is precisely known. Now we may calculate  the probability  $f_{\theta,\phi_0}(m_i)=\mathrm{Tr}(\rho_{\theta,\phi_0}\outpr{m_i}{m_i})$ corresponding to the eigenvalue $m_i$.
QFI is defined in terms of non-zero probability  distributions as \cite{Metro_QFI_review}
\begin{eqnarray}
F_\theta(\rho_{\theta_0,\phi_0},M)=\sum_{i, f\neq 0}\frac{1}{f_{\theta_0,\phi_0}(m_i)}\left(\at{\frac{\partial f_{\theta,\phi_0}(m_i)}{\partial\theta}}{\theta_0}\right)^2.~~~~~~~
\label{FrhothM}
\end{eqnarray}
Here $\at{\partial f_{\theta,\phi_0}(m_i)/\partial\theta}{\theta_0}$ quantifies the sensitivity of the observable $M$ to small fluctuations around $\theta_0$.  Similarly, if the polar angle is held fixed at $\theta_0$, while distributing azimuthal angle $\phi$ around $\phi_0$, QFI is then given by
\begin{eqnarray}
F_\phi(\rho_{\theta_0,\phi_0},M)=\sum_{i, f\neq 0}\frac{1}{f_{\theta_0,\phi_0}(m_i)}\left(\at{\frac{\partial f_{\theta_0,\phi}(m_i)}{\partial\phi}}{\phi_0}\right)^2. ~~~~~~
\label{FrhothMphi}
\end{eqnarray}
In the following we consider the specific cases of a single-qubit and an $N$-qubit STR and estimate the QFI corresponding to polar, azimuthal, and dual-parameters.

\subsubsection{QFI of a single-qubit: Polar parameter}\label{QFI_1qubit}
Consider a single target qubit prepared in the state 
\begin{eqnarray}
\varrho_{\theta,\phi_0} =\mathbbm{1}_2/2+\varepsilon_{t,1} \sigma_{\theta,\phi_0}/2,
\end{eqnarray}
where $\theta$ is in the neighborhood of $\theta_0$ (see Eq. (\ref{varrhoth0ph0})).

Since QFI depends on the observable $M$, it is natural to ask which observable maximizes QFI.  Such an optimal observable which maximizes QFI is known as the unbiased observable $M_{\overleftrightarrow{\theta_0},\phi_0}$ and it satisfies the flow equation
\begin{eqnarray}
\at{\frac{\partial\varrho_{\theta,\phi_0}}{\partial\theta}}{\theta_0}=\frac{1}{2}\left\{M_{\overleftrightarrow{\theta_0},\phi_0}\varrho_{\theta_0,\phi_0}+\varrho_{\theta_0,\phi_0} M_{\overleftrightarrow{\theta_0},\phi_0}\right\}.~~~
\label{L_eq_motion}
\end{eqnarray}
The solution of this equation leads to the unbiased observable in the form of a symmetric logarithmic derivative (SLD),
\begin{eqnarray}
M_{\overleftrightarrow{\theta_0},\phi_0} =\sum_{i,j,\lambda_i\neq-\lambda_j}\frac{2\left\langle\lambda_i \left\vert\at{\frac{\partial\varrho_{\theta,\phi_0}}{\partial\theta}}{\theta_0}  \right\vert \lambda_j \right \rangle} {\lambda_i+\lambda_j}\outpr{\lambda_i}{\lambda_j},~~~~~
\label{L_defined}
\end{eqnarray}
where $\lambda_i$ and $\ket{\lambda_i}$ are the eigenvalues and eigenvectors of $\varrho_{\theta_0,\phi_0}$ \cite{Metro_QFI_review}. Since the partial derivative
\begin{eqnarray}
\frac{\partial\varrho_{\theta,\phi_0}}{\partial\theta} = 
\frac{\varepsilon_{t,1}}{2}
\left[
\begin{array}{cc}
-\sin\theta & e^{-i\phi_0} \cos\theta   \\
e^{i\phi_0}\cos\theta  & \sin\theta
\end{array}
\right]=\frac{\varepsilon_{t,1}}{2}\frac{\partial\hat{n}}{\partial\theta}.\vec{\sigma},~~~~~
\label{rhothdot}
\end{eqnarray}
the corresponding unbiased observable for a single target qubit turns out to be (using Eqs. (\ref{sigthph}) and (\ref{L_defined}))  \cite{QFI_Bloch_sphere_PRA}
\begin{eqnarray}
M_{\overleftrightarrow{\theta_0},\phi_0} = 2 \at{\frac{\partial\varrho_{\theta,\phi_0}}{\partial\theta}}{\theta_0}=\varepsilon_{t,1}\at{\frac{\partial\hat{n}}{\partial\theta}}{\theta_0}.\vec{\sigma}.
\end{eqnarray}
Since $\hat{n}_0.\at{\frac{\partial\hat{n}}{\partial\theta}}{\theta_0}=0$, the unbiased observable corresponds to a direction orthogonal to the target state $\varrho_{\theta_0,\phi_0}$.

\begin{figure}
	\centering
	\includegraphics[width=9cm,clip = true,trim = 0cm 0.1cm 0cm 0cm]{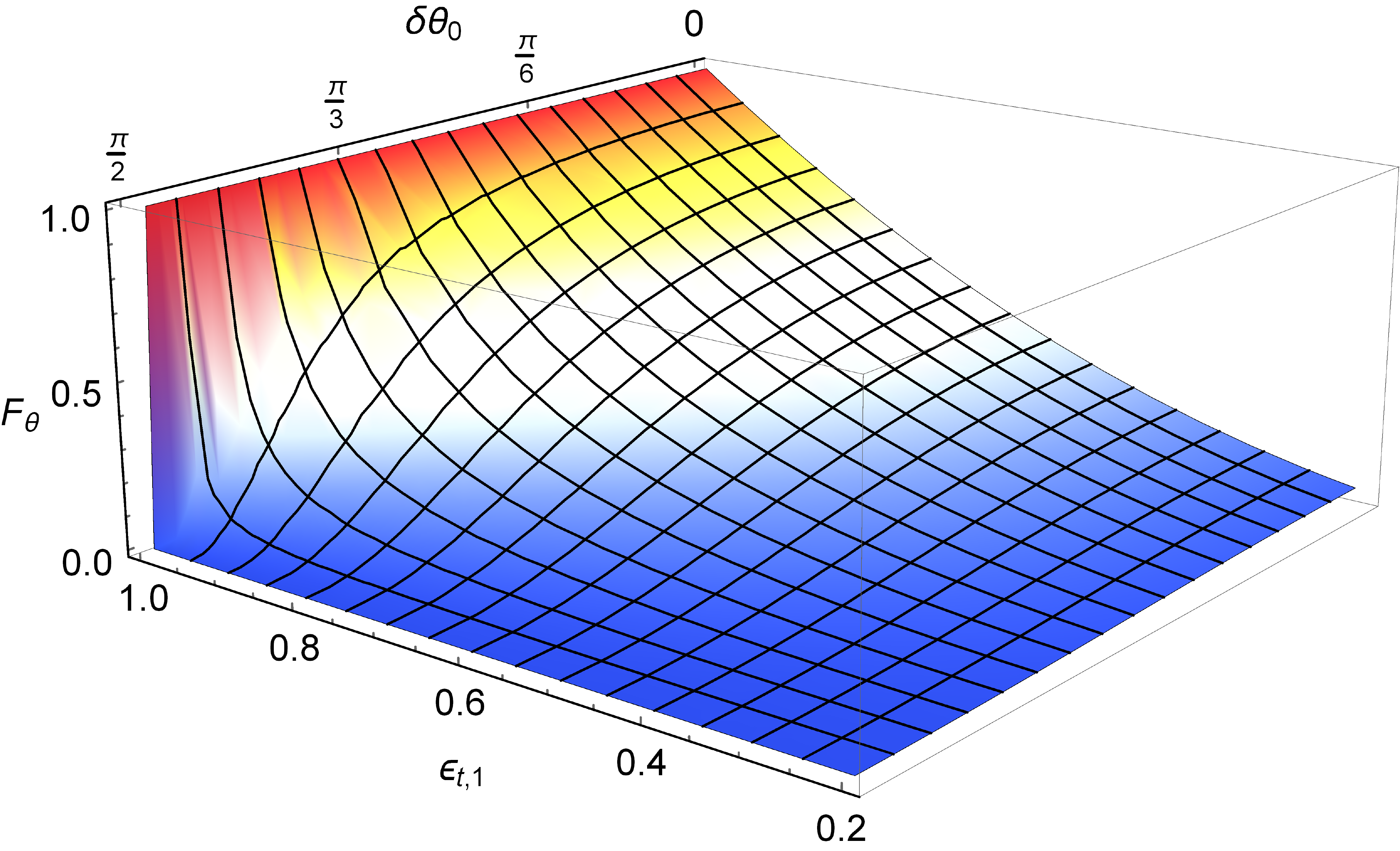}
	\caption{(Color online) Profile of QFI $F_\theta$ versus the deviation $\delta \theta_0$ in the polar angle and the purity $\varepsilon_{t,1}$ as described by Eq. (\ref{Fbiasedsinglequbit}).}
	\label{qfivsepth}
\end{figure}

Often the measurement observable is not the same as the optimal (unbiased) observable.  For example, a QST observable makes no prior assumption about the target state, and hence is in general a biased observable. To study QFI under a biased observable, we now consider a deviation of a chosen observable from the optimal observable via $\Theta_0 = \theta_0+\delta\theta_0$ and $\Phi_0 = \phi_0+\delta\phi_0$. The chosen (or  biased) observable is of the form
\begin{eqnarray}
M_{\overleftrightarrow{\Theta_0},\Phi_0} = 
\varepsilon_{t,1}
\left[
\begin{array}{cc}
	-\sin\Theta_0 & e^{-i\Phi_0} \cos\Theta_0   \\
	e^{i\Phi_0}\cos\Theta_0  & \sin\Theta_0
\end{array}
\right].
\end{eqnarray}
QFI obtained using Eq. (\ref{FrhothM})  is then
\begin{eqnarray}
 &&F_\theta(\varrho_{\theta_0,\phi_0},M_{\overleftrightarrow{\Theta_0},\Phi_0}) = \nonumber \\ &&\frac{\varepsilon_{t,1}^2(\cos \delta \phi_0 \cos\theta_0 \cos \Theta_0 + \sin \theta_0 \sin \Theta_0 )^2}{1-\varepsilon_{t,1}^2(\cos\delta \phi_0 \sin \theta_0 \cos\Theta_0 -\cos\theta_0 \sin \Theta_0)^2}. \nonumber
\end{eqnarray} 
For  $\delta \phi_0 = 0$, we obtain,
\begin{eqnarray}
F_\theta(\varrho_{\theta_0,\phi_0},M_{\overleftrightarrow{\Theta_0},\phi_0}) = \frac{\varepsilon_{t,1}^2 \cos^2 \delta \theta_0}{1-\varepsilon_{t,1}^2 \sin^2 \delta \theta_0}.
\label{Fbiasedsinglequbit}
\end{eqnarray}
Fig. \ref{qfivsepth} displays the profile of QFI in the above scenario. 
It can be observed that, for the optimal case of $\delta\phi_0 = 0$ and $\delta \theta_0 = 0$ (i.e., SLD), we obtain the upper bound for the mixed state QFI, i.e.,
\begin{eqnarray}
F_\theta(\varrho_{\theta_0,\phi_0},M_{\overleftrightarrow{\theta_0},\phi_0}) = \varepsilon_{t,1}^2=\mathrm{Tr}\left[\varrho_{\theta_0,\phi_0}\left\{ M_{\overleftrightarrow{\theta_0},\phi_0}\right\}^2\right],
\label{QFIsinglqubitSLD}~~~
\end{eqnarray}
since $\left\{ M_{\overleftrightarrow{\theta_0},\phi_0}\right\}^2 = \varepsilon_{t,1}^2 \mathbbm{1}_2$ \cite{QFI_Bloch_sphere_PRA}.
Also, for the maximally biased observable with $\delta \theta_0 = \pi/2$, QFI vanishes throughout. 

As a specific example, for the state   $\rho_{0,0} = \proj{0}$, we obtain $M_{\overleftrightarrow{0},0}= \sigma_x$ as the unbiased observable, and the maximum QFI, $F_\theta(\rho_{0,0},\sigma_x) = 1$.

An important application of QFI  is that it provides a bound to the variance $(\Delta \theta)^2$, via quantum Cramer-Rao bound
\begin{eqnarray}
(\Delta\theta)^2\ge \frac{1}{kF_{\theta}(\varrho_{\theta_0,\phi_0},M_{\overleftrightarrow{\theta_0},\phi_0})} = \frac{1}{k\varepsilon_{t,1}^2},
\end{eqnarray}
where $k$ is the number of independent measurements on identically prepared states in the neighborhood of $\varrho_{\theta_0,\phi_0}$ \cite{Metro_kolodyski_QFI}. In the NMR case, the number of independent measurements $k \sim 10^{15}$, same as the number of molecules in the experimental sample.  Taking $\varepsilon_{t,1}\sim 10^{-5}$, we obtain
$F_\theta \sim 10^{-10}$.  Nevertheless, 
$\Delta \theta < 10^{-2}$, which represents a reasonably high precision.

\subsubsection{QFI of a single-qubit: Azimuthal parameter}
Proceeding in a similar fashion as above, we obtain
\begin{eqnarray}
M_{\theta_0,\overleftrightarrow{\phi_0}} = 2 \at{\frac{\partial\varrho_{\theta_0,\phi}}{\partial\phi}}{\phi_0}=\varepsilon_{t,1}\at{\frac{\partial\hat{n}}{\partial\phi}}{\phi_0}.\vec{\sigma},
\end{eqnarray}
using Eq. (\ref{sigthph}).  Since $\hat{n}_0.\at{\frac{\partial\hat{n}}{\partial\phi}}{\phi_0}=0$, to achieve optimal measurement one has to measure in a direction orthogonal to the state $\varrho_{\theta_0,\phi_0}$.
Again, we consider a deviation of a chosen observable from the optimal observable via $\Theta_0 = \theta_0+\delta\theta_0$ and $\Phi_0 = \phi_0+\delta\phi_0$, and the corresponding biased observable is then 
\begin{eqnarray}
M_{\Theta_0,\overleftrightarrow{\Phi_0}} = 
\varepsilon_{t,1}\sin\Theta_0
\left[
\begin{array}{cc}
0 & -ie^{-i\Phi_0}   \\
ie^{i\Phi_0}  &0
\end{array}
\right].
\end{eqnarray}
QFI obtained using Eq. (\ref{FrhothMphi})  is then
\begin{eqnarray}
F_\phi(\varrho_{\theta_0,\phi_0},M_{\Theta_0,\overleftrightarrow{\Phi_0}}) =  \frac{\varepsilon_{t,1}^2\cos^2 \delta \phi_0 \sin^2\theta_0}{1-\varepsilon_{t,1}^2\sin^2 \delta \phi_0 \sin^2\theta_0}, \nonumber
\end{eqnarray} 
which is independent of $\delta\theta_0$. For the unbiased observable (SLD) i.e., $\delta\theta_0=\delta \phi_0 = 0$, we obtain,
\begin{eqnarray}
F_\phi(\varrho_{\theta_0,\phi_0},M_{\theta_0,\overleftrightarrow{\phi_0}})=\varepsilon_{t,1}^2\sin^2\theta_0=\mathrm{Tr}\left[\varrho_{\theta_0,\phi_0}\left\{ M_{\theta_0,\overleftrightarrow{\phi_0}} \right\}^2\right]\nonumber\\
\label{QFI_singlqubt_ph}
\end{eqnarray}
\cite{QFI_Bloch_sphere_PRA}. The quantum Cramer-Rao bound in this case is therefore
\begin{eqnarray}
(\Delta\phi)^2\ge \frac{1}{kF_\phi(\varrho_{\theta_0,\phi_0},M_{\theta_0,\overleftrightarrow{\phi_0}})} = \frac{1}{k\varepsilon_{t,1}^2\sin^2\theta_0}.
\end{eqnarray}

\subsubsection{QFI of a single-qubit: Dual parameters - $\theta$ and $\phi$}
In this case, we consider independent measurement of $\theta$ and $\phi$ and hence the corresponding QFIs are $F_\theta(\varrho_{\theta_0,\phi_0},M_{\overleftrightarrow{\theta_0},\phi_0})$ and $F_\phi(\varrho_{\theta_0,\phi_0},M_{\theta_0,\overleftrightarrow{\phi_0}})$. We now seek an effective dual parameter QFI, denoted by $\mathbbm{F}(\varrho_{\theta_0,\phi_0})$, which quantifies the maximum overall information.  To this end we utilize two-parameter quantum Cramer-Rao bound given by \cite{Metro_multi_parmtr_PRA}
\begin{eqnarray}
(\Delta \theta)^2+(\Delta \phi)^2 \ge \frac{1}{k\mathbbm{F}(\varrho_{\theta_0,\phi_0})}.
\label{QFIjoint_M_u0}
\end{eqnarray}
Here the effective dual-parameter QFI is related to the single-parameter QFIs via
\begin{eqnarray}
\frac{1}{\mathbbm{F}(\varrho_{\theta_0,\phi_0})} &=& \frac{1}{F_\theta(\varrho_{\theta_0,\phi_0},M_{\overleftrightarrow{\theta_0},\phi_0})}
	+ \frac{1}{F_\phi(\varrho_{\theta_0,\phi_0},M_{\theta_0,\overleftrightarrow{\phi_0}})} \nonumber
	\\  &=& \frac{1+\sin^2\theta_0}{\varepsilon_{t,1}^2\sin^2\theta_0}.
\label{QFIthph}
\end{eqnarray}

\subsubsection{QFI of an N-qubit STR: Polar parameter}\label{QFI_Nqubit}
Let us first consider an N-qubit STR prepared in a precorrelated initial state in the neighborhood of $\rho_{\theta_0,\phi_0}$ described in Eq. (\ref{rho2}).  In this case,
maximum QFI corresponding to an unbiased observable $M_{\overleftrightarrow{\theta_0},\phi_0}$ (SLD) is given by (see Eq. (\ref{QFIsinglqubitSLD}))
\begin{eqnarray}
F_\theta(\rho_{\theta_0,\phi_0},M_{\overleftrightarrow{\theta_0},\phi_0})=\mathrm{Tr}\left[\rho_{\theta_0,\phi_0} \left\{M_{\overleftrightarrow{\theta_0},\phi_0}\right\}^2\right]
\label{Fmax}
\end{eqnarray}
\cite{Metro_kolodyski_QFI}.
Using the form of $M_{\overleftrightarrow{\theta_0},\phi_0}$ as in Eq. (\ref{L_defined}), we obtain
\begin{eqnarray}
F_\theta(\rho_{\theta_0,\phi_0},M_{\overleftrightarrow{\theta_0},\phi_0})=\sum_{i,j,\lambda_i\neq-\lambda_j}\frac{4\left\vert\left\langle\lambda_i\left\vert\at{\frac{\partial\rho_{\theta,\phi_0}}{\partial\theta}}{\theta_0}\right\vert\lambda_j\right\rangle\right\vert^2}{(\lambda_i+\lambda_j)^2}\lambda_i.\nonumber\\
\label{FrhothMu}
\end{eqnarray}

Now we apply the following semi-analytical approach to analyze the above equation.  First we choose particular values for $N$ and $\varepsilon_{a,1}$ and a random value for $\phi_0$, to setup $\rho_{\theta,\phi_0}$ with arbitrary $\theta$ variable. This allows us to calculate the partial derivative $\partial \rho_{\theta,\phi_0}/\partial\theta$ and evaluate it at $\theta=\theta_0$. We diagonalize $\rho_{\theta_0,\phi_0}$ to obtain  eigenvalues $\lambda_i$ and corresponding eigenvectors $\ket{\lambda_i}$.  QFI can now be estimated using Eq. (\ref{FrhothMu}).  Finally we varied $N$ and studied the profile of $F_\theta(\rho_{\theta_0,\phi_0},M_{\overleftrightarrow{\theta_0},\phi_0})/\varepsilon_{a,1}^2$ over a randomized distribution of $\theta_0$ and $\phi_0$.
Using such a semi-analytical approach, we found that the maximum QFI has the form
\begin{eqnarray}
F_\theta(\rho_{\theta_0,\phi_0},M_{\overleftrightarrow{\theta_0},\phi_0})=\varepsilon_{a,1}^2(N-1),
\label{Fthph0}
\end{eqnarray}
where $N \ge 2$. Thus, in an STR with the target qubit precorrelated with ancillary qubits, all with small purity factors, QFI grows linearly with the number of ancillary qubits and quadratically with the purity factor $\varepsilon_{a,1}$ (for small purity, $\varepsilon_{a,1}\ll 1$).  Accordingly, the quantum Cramer-Rao bound for the variance $(\Delta \theta)^2$ in this case is
\begin{eqnarray}
(\Delta \theta)^2 \ge \frac{1}{k \varepsilon_{a,1}^2(N-1)}.
\end{eqnarray}
It can be noted that a similar precorrelation between probe and ancillary qubits in the presence of noise also leads to the enhancement in QFI \cite{Metro_precorltd_ancla_with_probe}.

However consider the uncorrelated state 
\begin{eqnarray}
	\rho^{\mathrm{uc}}_{\theta_0,\phi_0}=\mathbbm{1}_{2^N}/2^N+\varepsilon_{t,N}(\sigma_{\theta_0,\phi_0}\otimes \mathbbm{1}_{2^{N-1}})/2+\varepsilon_{a,N}\sum_{i=2}^N I_{iz}\nonumber\\
	\label{rho_th_uncorltd_NMR}
\end{eqnarray}
(see Eq. (\ref{rho0})). In this case, again using the semi-analytical approach, we found that
\begin{eqnarray}
F_\theta(\rho^{\mathrm{uc}}_{\theta_0,\phi_0},M_{\overleftrightarrow{\theta_0},\phi_0})\approx \varepsilon_{t,1}^2(1+F_\theta(\rho_{\theta_0,\phi_0},M_{\overleftrightarrow{\theta_0},\phi_0})) \approx \varepsilon_{t,1}^2,\nonumber
\end{eqnarray} 
since $F_\theta(\rho_{\theta_0,\phi_0},M_{\overleftrightarrow{\theta_0},\phi_0}) \ll 1$ in small purity states, which is no better than the single qubit case described in Eq. (\ref{QFIsinglqubitSLD}).
Thus ancillary qubits offer no advantage unless they are precorrelated with the target. This implies that $\rho^{\mathrm{uc}}_{\theta_0,\phi_0}$ is equivalent to $\varrho_{\theta_0,\phi_0}$ with respect to QFI.

\subsubsection{QFI of an N-qubit STR: Azimuthal parameter}
Again we consider an N-qubit STR prepared in the neighborhood of a precorrelated initial state described by  Eq. (\ref{rho2}).
Using similar methods applied for obtaining Eq. (\ref{Fthph0}), we found 
\begin{eqnarray}
F_\phi(\rho_{\theta_0,\phi_0},M_{\theta_0,\overleftrightarrow{\phi_0}}) = r\varepsilon_{a,1}^2(N-1),
\label{QFI_Mu_phi}
\end{eqnarray}
where the factor $r \in [0,1]$ depends on $\theta_0$ and $\phi_0$, and $N\ge 2$.
The corresponding quantum Cramer-Rao bound for the variance $(\Delta \phi)^2$ is
\begin{eqnarray}
(\Delta \phi)^2 \ge \frac{1}{kr\varepsilon_{a,1}^2(N-1)}.
\end{eqnarray}
However consider the uncorrelated state $\rho^\mathrm{uc}_{\theta_0,\phi_0}$
described in Eq. (\ref{rho_th_uncorltd_NMR}). In this case, using the semi-analytical approach described earlier, we found that
\begin{eqnarray}
F_\phi(\rho^\mathrm{uc}_{\theta_0,\phi_0},M_{\theta_0,\overleftrightarrow{\phi_0}})=\varepsilon_{t,1}^2\sin^2\theta_0.~~~~
\end{eqnarray} 
Comparing the above equation with Eq. (\ref{QFI_singlqubt_ph}), we find 
no advantage of ancillary qubits unless they are precorrelated with the target and that, $\varrho_{\theta_0,\phi_0}$ and
$\rho^\mathrm{uc}_{\theta_0,\phi_0}$ are equivalent with respect to QFI.

\subsubsection{QFI of an N-qubit STR: Dual parameters - $\theta$ and $\phi$}
Just like the one-qubit case (see Eqs. (\ref{QFIjoint_M_u0}) and (\ref{QFIthph})), the dual-parameter quantum Cramer-Rao bound in the N-qubit STR is given by \cite{Metro_multi_parmtr_PRA}
\begin{eqnarray}
(\Delta \theta)^2+(\Delta \phi)^2 \ge \frac{1}{k\mathbbm{F}(\rho_{\theta_0,\phi_0})},
\label{QFIjoint_M_u}
\end{eqnarray}
where 
\begin{eqnarray}
\mathbbm{F}(\rho_{\theta_0,\phi_0}) = \varepsilon_{a,1}^2(N-1)r/(1+r)
\label{strqfithph}
\end{eqnarray}
is the effective dual-parameter QFI.

\section{Experiments and Numerical Estimations}\label{Expt_sectn}
The experiments were carried out on a Bruker 500 MHz high-resolution NMR spectrometer using a liquid sample containing $300~\upmu$l of acetonitrile (H$_3$C$_2$N) dissolved in $400~\upmu$l of deuterated acetonitrile (D$_3$C$_2$N) at $300$ K. We used the spin-1/2 nuclei of naturally abundant  $^{13}$C nucleus as the target qubit and three spin-1/2 hydrogen nuclei of the methyl group as the ancillary qubits (qubits 2, 3, and 4) (see Fig. \ref{Pulse_seq_fig}(b)).
In this spin-system, the indirect spin-spin C-H couplings were  $J_{1i}=136.2$ Hz ($i=2,3,4$) while the H-H couplings are effectively nullified by magnetic equivalence.  We had chosen on-resonance carrier frequencies for both the nuclear species.  Various steps in the QST of a target qubit without and with ancillary qubits are described in Fig. \ref{Pulse_seq_fig}(c) and \ref{Pulse_seq_fig}(d) respectively.
In the following we describe estimation of QFI of the target qubit with and without exploiting the ancilla, and experimental QST with an STR.

 \begin{table*}
 	\begin{tabular}{|c|c|c|c|c|c|c|c|}
 		\hline   \multirow{5}{*}{$\sigma_{\theta_0,\phi_0}$} & \multirow{5}{*}{$C$}
 		&\multicolumn{6}{c|}{QFI} \\ \cline{3-8}
 		& &\multicolumn{3}{c}{With QST-based observables}& \multicolumn{3}{|c|}{With optimal observables (SLD)}\\\cline{3-8}
 		
 		&&&&&&& \\

 		& & $\mathbbm{F}_Q(\varrho_{\theta_0,\phi_0})/\varepsilon_{a,1}^2$ & $\mathbbm{F}_Q(\rho_{\theta_0,\phi_0})/\varepsilon_{a,1}^2$ & Amplification &$\mathbbm{F}(\varrho_{\theta_0,\phi_0})/\varepsilon_{a,1}^2$ &$\mathbbm{F}(\rho_{\theta_0,\phi_0})/\varepsilon_{a,1}^2$& Amplification\\

 		& & (Uncorrelated) & (Correlated STR) &  & (Uncorrelated)  &(Correlated STR)& \\

 		\hline  
 		$\sigma_{0,\phi_0}$ & 0.994 & 0& 0 & - & 0&0 &- \\ 
 		\hline  $\sigma_{\pi/2,0}$ & 0.984& 0.016&0.165 &10&0.031 &1.5&48 \\ 
 		\hline  $\sigma_{\pi/2,\pi/2}$ &0.998&0.016 & 0.186&12 &0.031 &1.5&48 \\ 
 		\hline  $\sigma_{\pi/4,0}$ &0.999 &0.008& 0.109 &14& 0.021 &1.0 &48\\ 
 		\hline  $\sigma_{\pi/4,\pi/2}$ & 0.999 &0.008& 0.149&19 &0.021 &1.0 &48\\ 
 		\hline
 	\end{tabular}
 	\caption{Experimental correlations ($C$) obtained with QST, and estimated QFIs for a set of states and corresponding QFI-amplification factors under various scenarios. Note that corresponding to $\theta_0=0$, the azimuthal parameter $\phi_0$ is indeterminate (hence dual parameter QFIs vanish). Hence the corresponding state is $\sigma_{0,\phi_0}$. }
 	\label{tab2}
 \end{table*}
 
 \subsubsection{Estimating QFI of an isolated (uncorrelated) qubit}\label{q1qfi}
 Here we estimate QFI of the target qubit for
 the following states:
 (i) $\sigma_{0,0}$, (ii)  $\sigma_{\pi/2,0}$, (iii)  $\sigma_{\pi/2,\pi/2}$, (iv)  $\sigma_{\pi/4,0}$, and (v)  $\sigma_{\pi/4,\pi/2}$.
 As explained in Sec. \ref{qst1q} and in Fig. \ref{Pulse_seq_fig}(c), the first step of QST involves the direct quadrature detection of $\varrho_{\theta_0,\phi_0}$ to determine $\phi_0$.  Since the quadrature detection involves partitioning the original signal into real and imaginary parts (using a reference wave with 0 and $\pi/2$ phase-shifts \cite{Levitt_Spindynabook}),
 we may use the additivity property of QFI \cite{QFIbookRoy} to obtain the effective azimuthal quadrature QFI 
\begin{eqnarray}
F_\phi(\varrho_{\theta_0,\phi_0},I_x+iI_y) = \frac{1}{2}F_\phi(\varrho_{\theta_0,\phi_0},I_x)+\frac{1}{2}F_\phi(\varrho_{\theta_0,\phi_0},I_y).\nonumber\\
\label{QFIaditvty_propNMR}
\end{eqnarray} 
Effective polar quadrature QFI $F_\theta(\varrho_{\theta_0,\phi_0},I_x+iI_y)$ is defined similarly. In Eq. (\ref{QFIaditvty_propNMR}), each term on the right hand side is estimated using Eq. (\ref{FrhothMphi}).
 However, the $\theta_0$ measurement involves destroying coherences using a pulsed field gradient followed by an $I_z$ measurement. 
 In NMR, $I_z$ measurement can be achieved by applying a $(\pi/2)_{y}$ pulse on the state followed by an $I_x$ measurement.  This allows us to estimate $F_\theta(\varrho_{\theta_0,\phi_0},I_z)=F_\theta(\varrho'_{\theta_0,\phi_0},I_x+iI_y)$ using expression for $F_\theta(\varrho_{\theta_0,\phi_0},I_x+iI_y)$ (see Eq. (\ref{QFIaditvty_propNMR})) and Eq. (\ref{FrhothM}).
 The dual parameter QFI corresponding to the QST observables is now obtained using the expression \cite{Metro_multi_parmtr_PRA}
 \begin{eqnarray}
 \frac{1}{\mathbbm{F}_Q(\varrho_{\theta_0,\phi_0})} = 
 \frac{1}{F_\theta(\varrho'_{\theta_0,\phi_0},I_x+iI_y)} + \frac{1}{F_\phi(\varrho_{\theta_0,\phi_0},{I_x+iI_y})}.\nonumber
 \label{QFIthph1}
 \end{eqnarray}
 
 Table \ref{tab2} lists the values of QFI $\mathbbm{F}_Q(\varrho_{\theta_0,\phi_0})$ for various initial states. However, if the target qubit is known to be in the neighborhood of $\varrho_{\theta_0,\phi_0}$ (QST do not require this information), one can perform optimal (unbiased) measurements to obtain $\mathbbm{F}(\varrho_{\theta_0,\phi_0})$ (as in Eq. (\ref{QFIthph})), whose estimated values are also listed in Table \ref{tab2}.

\subsubsection{A correlated target qubit in an STR: Experimental QST and estimation of QFI}
The experimental steps for preparing correlated STR, transformation of the target qubit into $\sigma_{\theta_0,\phi_0}$, QST, and measurements are described in Fig. \ref{Pulse_seq_fig}(d).
First we prepared a target-ancilla correlated state of the form $\rho_1$ (Eq. (\ref{rho1})). Then using a rotation of $\theta_0$ about a unit vector $\cos(\phi_0+\pi/2)\hat{x}+\sin(\phi_0+\pi/2)\hat{y}$ we rotate the target state $\sigma_{0,0}$ into $\sigma_{\theta_0,\phi_0}$ corresponding to each of the five states described in section \ref{q1qfi}.
As explained in section \ref{Tomo_with_precorltn} we experimentally performed QST of the target qubit ($^{13}$C) using a single-shot quadrature measurement of $M_Q$ observable (see Eq. (\ref{mqstr})) of the ancilla ($^1$H)  (with the help of tomography pulses; see $U_T$ in Fig. \ref{Pulse_seq_fig}(d)). 
Correlations (see Eq. (\ref{C_corelatn})) obtained for various states are tabulated in Table \ref{tab2}.
Similar to Eq. (\ref{QFIaditvty_propNMR}) we define the effective azimuthal quadrature QFI as
\begin{eqnarray}
F_\phi(\rho_{\theta_0,\phi_0},M_Q) = ~~~~~~~~~~~&& \nonumber\\ \frac{1}{2}\bigg\{ F_\phi(\rho_{\theta_0,\phi_0},\sum_qM_{qx}) 
 &+&  F_\phi(\rho_{\theta_0,\phi_0},\sum_qM_{qy})\bigg\}.\nonumber 
\label{QFIthXY}
\end{eqnarray}
The effective polar quadrature QFI $F_\theta(\rho_{\theta_0,\phi_0},M_Q)$ is defined in a similar way and the dual parameter QFI is estimated using the expression \cite{Metro_multi_parmtr_PRA}:
\begin{eqnarray}
\frac{1}{\mathbbm{F}_Q(\rho_{\theta_0,\phi_0})} = 
\frac{1}{F_\theta(\rho_{\theta_0,\phi_0},M_Q)} + \frac{1}{F_\phi(\rho_{\theta_0,\phi_0},M_Q)}.~~~~~~
\label{QFIthph2}
\end{eqnarray}

The estimated values of $\mathbbm{F}_Q(\rho_{\theta_0,\phi_0})$ are listed in table \ref{tab2}.
The estimated values of the QFI with optimal measurements i.e., $\mathbbm{F}(\rho_{\theta_0,\phi_0})$ (as in Eq. (\ref{strqfithph})) are also listed in Table \ref{tab2}.
We find that the QFI
$\mathbbm{F}_Q(\rho_{\theta_0,\phi_0})$
corresponding to the quadrature measurement of the correlated target qubit is amplified by an order of magnitude compared to the isolated (uncorrelated) qubit's QFI $\mathbbm{F}_Q(\varrho_{\theta_0,\phi_0})$.
Even the QFIs corresponding to the optimal measurements on the correlated target qubit are also amplified by a factor of 48 compared to that of the isolated qubit.  Interestingly, it can be related to the polarization enhancement
factor, which in the case of $N$-spin STR happens to be $(\varepsilon_{a,1}/\varepsilon_{t,1})\sqrt{N-1} = (\gamma_a/\gamma_t)\sqrt{N-1}$ \cite{NMR_Earnst_book}.  For acetonitrile this factor is about $6.93$.  Since QFI grows quadratically with the purity and linearly with number of ancillary qubits, one can expect $6.93^2 \simeq 48$ to be the amplification factor as evident from Table \ref{tab2}.
However $\mathbbm{F}_Q(\rho_{\theta_0,\phi_0})$ is much less than the maximum QFI corresponding to SLD i.e., $\mathbbm{F}(\rho_{\theta_0,\phi_0})$, this is because the former is obtained by QST-based observables with no prior information about the state of the target qubit, while the latter is obtained with optimized observables setup using the prior information that the target state is in the neighborhood of $\rho_{\theta_0,\phi_0}$. \\

\section{Summary and conclusion}\label{Sumry_conclusn}
Quantum Fisher information (QFI) is a tool to quantify the maximum achievable precision in measuring an unknown parameter with given amount of resource. We proposed and demonstrated in an NMR setup, a protocol to amplify QFI of a target qubit using a set of ancillary qubits.  For convenience, we chose a star topology register (STR), which consists of a central target qubit surrounded by a set of identical ancillary qubits. While an STR does not allow any individual control on the ancillary qubits, it allows the target qubit to efficiently correlate with all the ancillary qubits, leading to several interesting applications.
In this work, we showed that, if the target qubit is precorrelated with the ancillary qubits, it is possible to achieve a full quantum state tomography (QST) of the target qubit by a single quadrature measurement of only ancillary qubits.
We studied QFI of a target qubit that is correlated with the ancillary qubits and compared it with QFI of the uncorrelated target qubit.  In both cases, we estimated QFI corresponding to (i) the observables used for Quantum State Tomography (QST) with no prior information about the state of the target qubit and (ii) the optimal observables obtained given the state of the target qubit to be in the neighborhood of $\rho_{\theta_0,\phi_0}$. 
We showed that if the target qubit is initially precorrelated with ancillary qubits, we can achieve an order of magnitude amplification in QFI compared to the uncorrelated case even with QST observables. 
We further showed that with optimal observables, the amplification is not only higher, but also scales linearly with the size of the STR (i.e., number of ancillary qubits) and quadratically with the purity of individual ancillary qubits (for small purity, $\varepsilon_{a,1} \ll 1$).  We believe that this protocol is a step towards realizing efficient quantum measurements applicable for a variety of quantum architectures including spin-based architectures.\\

\section*{Acknowledgment}
We thank Abhishek Shukla for  suggestions which ultimately culminated in this work. We also acknowledge useful discussions with  Deepak Khurana and Anjusha V S. This work was supported by DST/SJF/PSA-03/2012-13 and CSIR 03(1345)/16/EMR-II.

\bibliographystyle{apsrev4-1}
\bibliography{bib_ch}

\end{document}